\documentclass[11pt]{article}
\usepackage{graphicx}
\advance\textheight by \topskip
\begin{document}
\tolerance=100000
\thispagestyle{empty}
\setcounter{page}{0}
\def\cO#1{{\cal{O}}\left(#1\right)}
\newcommand{\cp}{\mbox{$\not \hspace{-0.15cm} C\!\!P \hspace{0.1cm}$}}
\newcommand{\be}{\begin{equation}}
\newcommand{\ee}{\end{equation}}
\newcommand{\br}{\begin{eqnarray}}
\newcommand{\er}{\end{eqnarray}}
\newcommand{\ba}{\begin{array}}
\newcommand{\ea}{\end{array}}
\newcommand{\bi}{\begin{itemize}}
\newcommand{\ei}{\end{itemize}}
\newcommand{\bn}{\begin{enumerate}}
\newcommand{\en}{\end{enumerate}}
\newcommand{\bc}{\begin{center}}
\newcommand{\ec}{\end{center}}
\newcommand{\ul}{\underline}
\newcommand{\ol}{\overline}
\newcommand{\ra}{\rightarrow}
\newcommand{\sm}{${\cal {SM}}$}
\newcommand{\as}{\alpha_s}
\newcommand{\aem}{\alpha_{em}}
\newcommand{\ycut}{y_{\mathrm{cut}}}
\newcommand{\susy}{{{SUSY}}}
\newcommand{\Dir}{\kern -6.4pt\Big{/}}
\newcommand{\Dirin}{\kern -10.4pt\Big{/}\kern 4.4pt}
\newcommand{\DDir}{\kern -10.6pt\Big{/}}
\newcommand{\DGir}{\kern -6.0pt\Big{/}}
\def\Ecm{\ifmmode{E_{\mathrm{cm}}}\else{$E_{\mathrm{cm}}$}\fi}
\def\gluino{\ifmmode{\mathaccent"7E g}\else{$\mathaccent"7E g$}\fi}
\def\photino{\ifmmode{\mathaccent"7E \gamma}\else{$\mathaccent"7E \gamma$}\fi}
\def\mgluino{\ifmmode{m_{\mathaccent"7E g}}
             \else{$m_{\mathaccent"7E g}$}\fi}
\def\taugluino{\ifmmode{\tau_{\mathaccent"7E g}}
             \else{$\tau_{\mathaccent"7E g}$}\fi}
\def\mphotino{\ifmmode{m_{\mathaccent"7E \gamma}}
             \else{$m_{\mathaccent"7E \gamma}$}\fi}
\def\ML{\ifmmode{{\mathaccent"7E M}_L}
             \else{${\mathaccent"7E M}_L$}\fi}
\def\MR{\ifmmode{{\mathaccent"7E M}_R}
             \else{${\mathaccent"7E M}_R$}\fi}
\def\lsim{\buildrel{\scriptscriptstyle <}\over{\scriptscriptstyle\sim}}
\def\gsim{\buildrel{\scriptscriptstyle >}\over{\scriptscriptstyle\sim}}
\def\MCH {$\tilde\chi_1^+$}
\def \CH{{\tilde\chi}^{\pm}}
\def \LSP{\tilde\chi_1^0}
\def \SNU{\tilde{\nu}}
\def \BARSNU{\tilde{\bar{\nu}}}
\def \MLSP{m_{{\tilde\chi_1}^0}}
\def \MCH{m_{{\tilde\chi}^{\pm}}}
\def \MCHMIN {\MCH^{min}}
\def \ET{\not\!\!{E_T}}
\def \LL{\tilde{l}_L}
\def \LR{\tilde{l}_R}
\def \MLL{m_{\tilde{l}_L}}
\def \MLR{m_{\tilde{l}_R}}
\def \MSNU{m_{\tilde{\nu}}}
\def \PI{{\pi^{\pm}}}
\def \DM{{\Delta{m}}}
\newcommand{\bQ}{\overline{Q}}
\newcommand{\ad}{\dot{\alpha }}
\newcommand{\bd}{\dot{\beta }}
\newcommand{\dd}{\dot{\delta }}
\def \CH{{\tilde\chi}^{\pm}}
\def \MCH{m_{{\tilde\chi}_1^{\pm}}}
\def \LSP{\tilde\chi_1^0}
\def \MUL{m_{\tilde{u}_L}}
\def \MUR{m_{\tilde{u}_R}}
\def \MDL{m_{\tilde{d}_L}}
\def \MDR{m_{\tilde{d}_R}}
\def \MSNU{m_{\tilde{\nu}}}
\def \MLL{m_{\tilde{l}_L}}
\def \MLR{m_{\tilde{l}_R}}
\def \mhf{m_{1/2}}
\def \MST{m_{\tilde t_1}}
\def \lum{{\cal L}}
\def \RPVC{\lambda'}
\def\tth{\tilde{t}\tilde{t}h}
\def\qqh{\tilde{q}_i \tilde{q}_i h}
\def\t1{\tilde t_1}
\def \pt{p{\!\!\!/}_T} 
\def \etm{E{\!\!\!/}_T} 
\def\beq{\begin{eqnarray}}
\def\enq{\end{eqnarray}}

\def\lapp{\mathrel{\rlap{\raise.5ex\hbox{$<$}}
                    {\lower.5ex\hbox{$\sim$}}}}
\def\gapp{\mathrel{\rlap{\raise.5ex\hbox{$>$}}
                    {\lower.5ex\hbox{$\sim$}}}}
\newcommand{\decay}[2]{
\begin{picture}(25,20)(-3,3)
\put(0,-20){\line(0,1){10}}
\put(0,-20){\vector(1,0){15}}
\put(0,0){\makebox(0,0)[lb]{\ensuremath{#1}}}
\put(25,-20){\makebox(0,0)[lc]{\ensuremath{#2}}}
\end{picture}}
\vspace*{\fill}
\begin{flushright}
{IISc-CHEP/15/04}\\
{CERN-TH/2004-250}\\
{WUE-ITP-2004-037}\\
{hep-ph/0412193}\\
\end{flushright}
\begin{center}
{\Large \bf
Probing the CP-violating light neutral Higgs in the charged Higgs 
decay at the LHC \\[0.25cm]}
\end{center}
\begin{center}
{\large Dilip Kumar Ghosh$^{a}$, R.M. Godbole${^{b}}$ 
and D.P. Roy$^{c,d}$ }\\[0.25 cm]
$^a$ Institut f{\"u}r Theoretische Physik und Astrophysik, \\
Universit{\"a}t W{\"u}rzburg, D-97074, 
W{\"u}rzburg, Germany.\\[0.20cm]
$^b$Centre for High Energy Physics, Indian Institute of Science, Bangalore, 560 012, India.\\[0.20cm]
$^c$Department of Theoretical  Physics, Tata Institute of 
Fundamental Research,\\ Homi Bhabha Road, 400 005 Mumbai, India.\\[0.20cm]
$^d$ Theory Division, Physics Department, CERN, CH-1211 Geneva 23,
Switzerland.\\[0.20cm]
\end{center}

\vspace{.4cm}

\begin{abstract}
{\noindent\normalsize 
The CP-violating MSSM allows existence of a light neutral Higgs boson ($M_{H_1}
\lapp 50$ GeV) in the CPX scenario in the low $\tan \beta (\lapp 5)$ region,
which could have escaped the LEP searches due to a strongly suppressed 
$H_1 Z Z$  coupling. This parameter space corresponds to a relatively light 
$H^+$ ($M_{H^+} < M_t$) , which is predicted to decay dominantly into 
the $W H_1$ 
channel. Thus one expects to see a striking $t \bar t$ signal at the LHC, 
where one of the top quarks decays into the $bb \bar b W$ channel, via
$t \to b H^\pm, H^\pm \to W H_1$ and $H_1 \to b \bar b$. The characteristic 
correlation between the $b \bar b$, $b \bar b W$ and $b b \bar b W$ invariant
mass peaks is expected to make this signal practically free of the SM 
background. Our parton level Monte Carlo simulation yields upto 5000 events,
for ${\cal L} = 30$ fb$^{-1}$, over the parameter space of interest, after 
taking into account the b-tagging efficiency for three or more b-tagged jets.
}
\end{abstract}
\vskip1.0cm
\noindent
\vspace*{\fill}
\newpage
\section*{Introduction}

The search for Higgs bosons and  study of their properties is one of 
the main goals of physics studies at the Tevatron upgrade (Run 2) and the
upcoming Large Hadron Collider (LHC).  The precision measurements with 
Electro-Weak (EW) data indicate the existence of a light Higgs boson 
($M_h <246$~GeV at 95\% C.L.) whereas direct searches rule out the case 
$M_h < 114.4$ GeV \cite{Eidelman:2004wy}~\cite{lep}.
Naturalness arguments along with the indication of a light Higgs 
state suggest that Supersymmetry (SUSY) is a likely candidate for new physics 
Beyond the Standard Model (BSM). Even in the SUSY case, a mass for the 
lightest neutral Higgs smaller than $\sim 90$ GeV is ruled out~\cite{susylim} 
if the SUSY parameters as well as the SUSY breaking parameters are real and 
CP is conserved. However, in presence of CP-violation in the Higgs sector,
the lower limit can get diluted due to a reduction in the  $H_1 ZZ$ 
coupling~\cite{gunion}. 

CP violation, initially observed only in the
$K^0$--$\bar {K^0}$ system, is one feature of the Standard Model
(SM) that still defies clear theoretical understanding.
It is in fact one of the necessary ingredients for
generating the observed excess of baryons over antibaryons in the
Universe \cite{Sakharov:dj,Dolgov:2002kw}. The amount of CP
violation present in the quark sector described very satisfactorily in
the CKM picture, is however, too small to generate a baryon asymmetry of
the observed level of $N_B /N_{\gamma} \simeq 6.1 \times
10^{-10}$~\cite{Bennett:2003bz}. New sources of CP violation {\it beyond}
the SM are therefore a necessity~\cite{Dine:2003ax}.

The Minimal Supersymmetric
Standard Model(MSSM), in principle, admits a large number of phases which
can not be rotated away by a simple redefinition of the fields and hence 
provide new sources of CP-violation. A large number of these phases
involving the first two generations of sparticles are strongly
constrained by the electric dipole moments of the electron and neutron
(EDMs)\cite{edm1, edm2} and mercury atoms \cite{edm3}. However, these
constraints are model-dependent. It has been demonstrated
that cancellations among different diagrams allow certain
combinations of these phases to be large in a general MSSM.
Furthermore, if the sfermions of the first two generations are
sufficiently heavy, above the 1 TeV range, the EDM constraints on
the phase of the higgsino mass parameter $\mu=|\mu|e^{i\phi_\mu}$,
in general constrained to $\phi_\mu^{}\lsim 10^{-2}$, get weaker;
the sfermions of the third generation can still be light.

In a version of MSSM where the
Higgsino mass term $\mu$, the gaugino masses $M_i$ and the trilinear couplings
$A_f$ are complex the Higgs sector, even with CP-conserving tree level
scalar potential, has loop induced CP-violation~\cite{aposto1,aposto2,demir,
jsik,aposto3,kane-wang}. The LEP data can allow a much lighter Higgs with a mass
$\lapp 40$--$50$ GeV \cite{carena,Abbiendi:2004ww,susylim} due to a 
reduction in the $H_1 ZZ$ coupling in the CPX scenario~\cite{aposto2}, which  
corresponds to a certain choice of the CP-violating SUSY parameters, chosen so
as to showcase the CP-violation in the Higgs sector in this case.
In a large portion of this region all the usual search channels of such a light 
Higgs at the LHC are also not expected to be viable~\cite{carena} due to the
simultaneous reduction in the coupling of the Higgs to a vector boson pair as 
well as the $t\bar t $ pair. As a matter of fact presence of CP-violation in
Supersymmetry and hence the Higgs sector, can affect the Higgs decays as well 
as their production rates at the colliders substantially and has been a 
subject of many investigations~\cite{cpv-susyhiggs,carena}.

It is interesting to note that in the same region of the parameter space where 
the coupling of the lightest mass eigenstate $H_1$ to a pair of Z-bosons:  
the $H_1ZZ$ coupling, is suppressed the $H^+ W^- H_1$ coupling is enhanced 
because these two sets of couplings satisfy a sum-rule.  The strong suppression 
of the $H_1 ZZ$ coupling also means that the $H_1$ is dominated by the 
pseudo-scalar component in this region and hence implies a light charged 
Higgs boson ($M_{H^+} < M_t$). These two features suggest that $H^\pm 
\rightarrow H_1 W^\pm$ is the dominant decay mode of the $H^\pm$ over the 
parameter space of interest.  This motivated us to study the possibility of 
probing at the LHC, such a light Higgs scenario in CP violating MSSM Higgs model
through the process 
$p p \to t \bar t X \to (b W^\pm) (b H^\mp)  X \to (b \ell \nu)
(b H_1 W) X  \to (b \ell \nu) (b b \bar b ) (jj) + X  $
along with the hadronic and leptonic decays of the two $W'$s interchanged.
Thus the signal will consist of three or more b-tagged jets and two untagged 
jets along with a hard lepton and missing $p_T$. Similar studies have been 
done in the context of charged Higgs search in NMSSM model~\cite{dpr}.
In the next section we present the notation and some details of the 
calculation, followed by presentation of the results in the section after
that and we end by making some concluding remarks.

\section*{Notation and Formalism}

As already mentioned in the introduction the  non-vanishing phases of 
$\mu$ and/or the trilinear scalar couplings $A_{t,b}$ can induce explicit 
CP-violation in the Higgs sector via loop corrections. Thus the Higgs 
potential, even though invariant under 
CP-transformation at tree level, receives CP-violating contributions 
on loop corrections.  Due to large Yukawa interactions of the Higgs bosons 
to top and bottom squarks, ${\rm Arg}(\mu)$ and  ${\rm Arg}(A_t), 
{\rm Arg}(A_b)$ are the 
relevant CP phases.  These  generate contributions to the off diagonal 
block ${\cal M}^2_{\rm SP}$ in the $3\times 3$ neutral Higgs boson 
mass-squared matrix ${\cal M}^2_{ij}$,  mixing the scalar $(S)$ and the
pseudo-scalar $(P)$ Higgs fields \cite {aposto1,aposto2,demir,jsik,
aposto3,kane-wang}. These may be given approximately by \cite{aposto2}:
\beq
{\cal M}^2_{\rm SP} \approx {\cal O}\left 
( \frac{M^4_t  \mid \mu \mid \mid A_t \mid}{v^2 32 \pi^2 M^2_{\rm SUSY}}\right )
\sin \Phi_{\rm CP} \nonumber \\ \times \left [6, \frac{\mid A_t \mid^2 }{M^2_{\rm SUSY}}, 
\frac{\mid \mu\mid^2}{\tan\beta M^2_{\rm SUSY}}, 
\frac{\sin 2\Phi_{\rm CP}\mid A_t\mid \mid\mu\mid }{\sin \Phi_{\rm CP} 
M^2_{\rm SUSY}}\right ]
\enq
where $\Phi_{\rm CP} = {\rm Arg}(A_t\mu)$, $v = 246$ GeV. From the above 
expression it is clear that sizeable scalar-pseudo-scalar mixing is 
possible for large CP violating phase $\Phi_{\rm CP}$,
$\mid \mu \mid $ and $\mid A_t \mid $ $(> M_{\rm SUSY})$. The mass scale
$M_{\rm SUSY}$ is defined by $(m^2_{\tilde t_1} + m^2_{\tilde t_2})/2 $.
After diagonalizing the $3\times 3$ symmetric Higgs mass-squared matrix 
${\cal M}^2_{ij}$ by an orthogonal matrix $O$, the physical mass 
eigenstates $H_1, H_2 $ and $H_3$ (in ascending order of mass) are  
states of indefinite  CP parity.  In this case $M_{H^\pm}$ is
more appropriate parameter for description of the MSSM Higgs-sector in place 
of the  $M_A$ used usually in the CP-conserving case.

As a result of the CP-mixing in the neutral Higgs sector, their couplings to
the gauge bosons and the fermions get modified.  For the purpose of 
illustration we provide the couplings of $H_iVV$, $H_iH_j Z$ and
$H_iH^\pm W^\mp $ below. More  details can be found in Ref. \cite{aposto2}.
\beq
{\cal L}_{H_iVV}&=& g M_W\ \sum_{i=1}^3 g_{H_i VV}
\ [H_iW_\mu^+W^{-,\mu} +\frac{1}{2 c_W^2} H_iZ_\mu Z^{\mu} ] \\
{\cal L}_{H_iH_j Z} &=& \frac{g}{2 c_W}\ \sum_{j>i=1}^3 
g_{H_iH_jZ}\ ( H_i\, \!\!
\stackrel{\leftrightarrow}{\vspace{2pt}\partial}_{\!\mu} H_j )Z^\mu \\
{\cal L}_{H H^\mp W^\pm}&=& \frac{g}{2 c_W}\ \sum_{i=1}^3
g_{H_iH^-W^+}\ ( H_i\, \!\!
\stackrel{\leftrightarrow}{\vspace{2pt}\partial}_{\!\mu} H^-)W^{+,\mu}
\enq
where,  $g_{H_iVV} $, $g_{H_iH_jZ}$ 
and $ g_{H_iH^+W^-}$ are Higgs gauge boson couplings
normalized to the standard model value and can be written as,
\beq
g_{H_iVV} &=& O_{1i}\cos\beta+ O_{2i}\sin\beta, \\
g_{H_iH_j Z} &= &O_{3i}(\cos\beta O_{2j}- \sin\beta O_{1j})
-(i\leftrightarrow j)\\
g_{H_iH^+W^-} &=& O_{2i}\cos\beta- O_{1i}\sin\beta + i O_{3i}
\enq
These couplings obey the following sum rules:
\beq
\sum_{i=1}^3 g^2_{H_i VV} & = & 1,\\
g^2_{H_i VV} + \mid g_{H_i H^+W^-}\mid^2 & = & 1, \\
g_{H_k VV}  & = & \epsilon_{ijk} g_{H_i H_j Z} 
\end{eqnarray}
From the above sum rules one can see that if two of the  $g_{H_i ZZ}$ are known,
then the whole set of couplings of the neutral Higgs boson to the gauge
bosons are determined. It is interesting to see from Eq. (9)  that in
the presence of large CP-violating effects, with large scalar-pseudo-scalar
mixing, the suppressed $H_1 VV$ coupling means an enhanced $H_1 H^+W^-$ 
coupling. This enhancement  will play a significant role
in our analysis.  Equally important is the correlation between the 
mass of the charged Higgs $M_{H^\pm}$ and that of the pseudo-scalar state that 
exists in the MSSM.  A suppressed $H_1 VV$ coupling implies a light 
pseudo-scalar state,  which  in turn implies a light charged Higgs, 
with $M_{H^+} < M_t$.  

As has been discussed before, the quantity 
$\sin \Phi_{\rm CP}/M^2_{\rm SUSY}$ needs to be large  to get significant
CP-mixing in the Higgs sector.  The CP-{\it violating benchmark scenario}
(CPX) has been suggested~\cite{aposto2} to showcase this CP-violation  and
provides a  suitable set of parameters which can be used to study the 
phenomenology of the CP-violating MSSM Higgs sector:
\beq
\widetilde{M}_Q=\widetilde{M}_t=\widetilde{M}_b=M_{\rm SUSY}\\
\mu=4 M_{\rm SUSY}, |A_t|=|A_b|= 2 M_{\rm SUSY},\\
{\rm Arg}(A_t)={\rm Arg}(A_b)
\enq
In the next section we first summarize the current constraints from LEP on the
MSSM parameter space and hence on the Higgs masses in the CPX scenario and then
discuss the phenomenology of the charged and the neutral Higgs search in the 
region of  the low $M_{H_1}$ window that is still allowed by 
LEP~\cite{carena,Abbiendi:2004ww,susylim} for the case of CP-violating MSSM.

\section*{Results and discussion}

Recently the OPAL Collaboration \cite{Abbiendi:2004ww} reported 
their results for the Higgs boson searches in the CP-violating MSSM Higgs 
sector using the parameters defined in the CPX scenario as mentioned above 
and found that for certain values of phases and $M_{H^+}$, the lower
mass limit on the neutral Higgs is diluted, at times vanishing completely. 
This results in windows in the $\tan \beta$--$M_{H^+}$ plane which are
still allowed by the LEP data. The LEP bounds are essentially evaded in 
this window as the lightest state is largely a pseudo-scalar with highly 
suppressed coupling to the $ZZ$ pair. There exist two programs; 
CPSuperH~\cite{cpsuperh} and  FeynHiggs 2.0~\cite{Heinemeyer:2001qd} 
to calculate the masses and mixing in the Higgs sector in the CP-violating case.
Due to the different approximations made in the
two calculations as well as differences in the inclusion of different higher
order terms, at least in the CPX scenario, the two programs give somewhat 
different results and the experimentalists use the lower prediction
of the two for the expected cross-sections to get the most conservative 
constraints. The constraints also depend sensitively on the mass of the top
quark used in the calculation~\cite{susylim}.  The preliminary results from
a combined analysis of all the LEP results~\cite{susylim}, provide exclusion
regions in the $M_{H_1}- \tan \beta $ plane for different values of the 
CP-violating phases, for the following values of the parameters:
\beq
{\rm Arg}A_t = {\rm Arg}A_b = {\rm Arg}M_{\tilde g}=\Phi_{\rm CP},\\
M_{\rm SUSY}= 0.5~{\rm TeV}, M_{\tilde g} = 1~ {\rm TeV},\\  
M_{\tilde B} = M_{\tilde W}= 0.2~ {\rm TeV},\\
\Phi_{\rm CP} = 0^{\circ}, 30^{\circ}, 60^{\circ}, 90^{\circ}.
\enq
Combining the results of Higgs searches from ALEPH, DELPHI, L3 and OPAL,
the authors in Ref.\cite{carena} have also provided exclusion regions in the
$M_{H_1}$--$\tan \beta $ plane as well as $M_{H^+}$--$\tan \beta $ plane for 
the same  set of parameters. 
\begin{table}
\label{phasesixty}
\begin{footnotesize}
\begin{center}
\begin{tabular}{|c|c|c|c|c|}
\hline
&&&&\\
$\tan\beta $ &$2$&2.2&2.5&3.0\\[3mm]
\hline
&&&&\\
${\rm Br} (H^+ \to H_1 W^+)(\%) $ & $>90$ (83.5) &$> 90 $(80.32)
&$>90$ $(73.85)$ &$>90 $ (63.95)  \\[3mm]
${\rm Br} (t \to b H^+)(\%) $ & 4.0 -- 4.2 & 4.9 -- 5.1
& 4.8 -- 5.11 & 4.0 -- 4.3 \\[3mm]
$M_{H^+}$ (GeV) & $< 133.6$ (135.1)&$< 122.7$ (124.3) &$< 113.8$
$(115.9)$
&$< 106.6$(109.7)\\[3mm]
$M_{H_1}$ (GeV) & $< 50.97$ (54.58) &$< 39.0$ (43.75) &$< 27.97$ (35.44)
&$< 14.28$ (29.21) \\[3mm]
\hline
\end{tabular}
\caption{Range of values  for BR ($H^+ \rightarrow H_1 W^+$) and
BR ($t \rightarrow b H^+ $)
for different values of $\tan \beta$ corresponding to the LEP allowed
window in the CPX scenario, for the common phase $\Phi_{\rm CP} = 60^{\circ}$,
along with the corresponding range for the $H_1$ and
$H^+$ masses. The quantities in the bracket in each column give the values
at the edge of the kinematic region where the decay
$H^+ \rightarrow H_1 W^+$ is allowed.}
\end{center}
\end{footnotesize}
\end{table}
While the exact exclusion regions differ somewhat in the three analyses
\cite{carena,Abbiendi:2004ww,susylim} they
all  show that for phases $\Phi_{\rm CP} = 90^{\circ} $ and $60^{\circ}$ 
LEP cannot exclude the presence of a light Higgs boson at low $\tan \beta$,
mainly because of the suppressed $H_1 ZZ$ coupling. The analysis of 
Ref.~\cite{carena} further shows that in the same region the $H_1 t \bar t$ 
coupling is suppressed as well.  Thus   this particular region in the
parameter space can not be probed either at the Tevatron where  the
associated production $W/Z H_1$ mode is the most promising one; neither can 
this be probed at the LHC as the reduced $t \bar t H_1$ coupling suppresses
the inclusive production mode and  the associated production modes
$W/Z H_1$ and $t \bar t H_1$, are suppressed as well. This region of
Ref.~\cite{carena} corresponds to $\tan \beta \sim 3.5-5, 
M_{H^+}\sim 125-140 $ GeV,~ $ M_{H_1} \stackrel{<}{{}_\sim} 50 $ GeV 
and $\tan \beta \sim 2-3, M_{H^+}\sim 105-130~{\rm GeV}, M_{H_1} 
\stackrel{<}{{}_\sim} 40 $ GeV, for $\Phi_{CP} = 90^\circ$ and $60^\circ$ 
respectively. The code CPSuperH  and $M_t = 175$ GeV has been used by them 
to calculate the couplings and the masses of the Higgs-bosons.

As mentioned already, in the same region of the parameter space where $H_1 ZZ$
coupling is suppressed, the $H^+ W^- H_1$ coupling is enhanced because these
two sets of couplings satisfy a sum-rule as shown in Eq.(9). Further, in the
MSSM a light pseudo-scalar implies a light charged Higgs, lighter than the 
top quark.
\begin{table}
\label{phaseninty}
\begin{footnotesize}
\begin{center}
\begin{tabular}{|c|c|c|c|c|}
\hline
&&&&\\
$\tan\beta $ &$3.6$&4&4.6&5\\[3mm]
\hline
&&&&\\
${\rm Br} (H^+ \to H_1 W^+)(\%) $ & $>90$(87.45)&$> 90 $(57.65) 
&$>90$ (50.95) &$>90$(46.57)  \\[3mm]
${\rm Br} (t \to b H^+)(\%) $ & $\sim$ 0.7 & .7 -- 1.1  
& 0.9 -- 1.3 & 1.0 -- 1.3  \\[3mm]
$M_{H^+}$ (GeV) & $< 148.5$ (149.9) &$< 139$ (145.8) &$< 130.1 $
(137.5) &$< 126.2$(134) \\[3mm]
$M_{H_1}$ (GeV) & $< 60.62$ (63.56)  &$< 49.51$ (65.4)  &$<36.62 $
(57.01) &$< 29.78$(53.49) \\[3mm]
\hline
\end{tabular}
\caption{
Same as in Table 1 but  for the value of common phase
$\Phi_{\rm CP} = 90^{\circ}$. }
\end{center}
\end{footnotesize}
\end{table}
\begin{figure}[hbt]
\label{fig1}
\includegraphics*[scale=0.6] {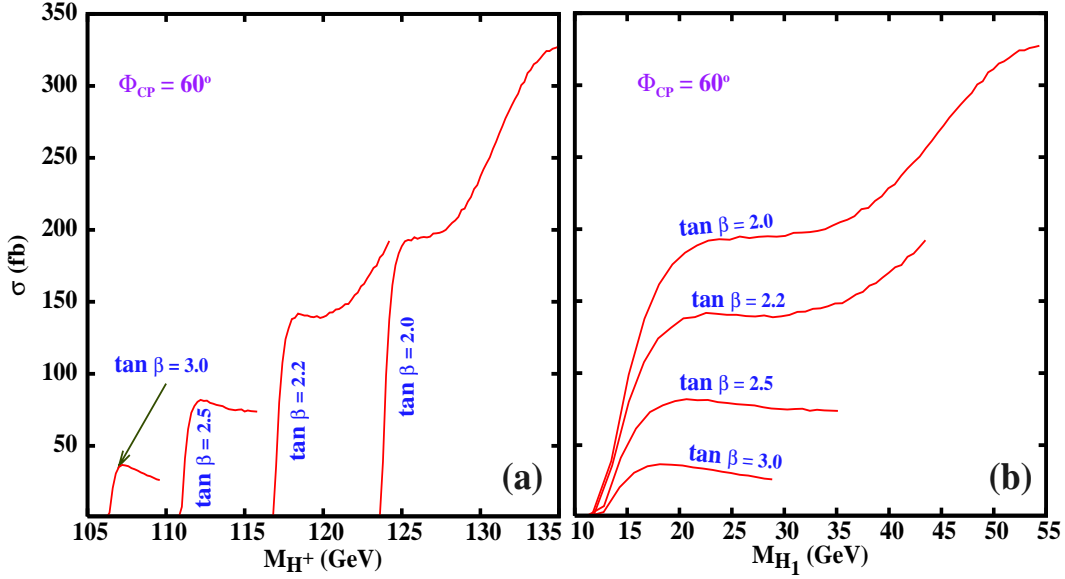}
\caption{Variation of the expected  cross-section with $M_{H^+}$ $(a)$ and
$M_{H_1}$ ($b)$ for four values of $\tan\beta =2, 2.2, 2.5 $ and $3$.
The CP-violating phase $\Phi_{\rm CP}$ is $60^{\circ}$. See text for the values
of the remaining  MSSM parameters. The cross-sections are obtained after 
applying the mass window cuts as mentioned in the text. These numbers 
should be multiplied by $\sim 0.5$ to get the signal cross-section as 
explained in the text.}
\end{figure}
Tables 1  and 2 show the behaviour of the $M_{H^+}$, $M_{H_1}$ and 
the BR ($H^+ \rightarrow H_1 W^+$), for values of $\tan \beta $ 
corresponding to the above mentioned window in the $\tan \beta$--$M_{H_1}$ 
plane,
of Ref.~\cite{carena}. It is to be noted here that indeed the $H^\pm$ is light 
(lighter than the top) over the entire range, making its production in $t$ 
decay possible. Further, the $H^\pm$ decays dominantly into $H_1W$, with a 
branching ratio larger than $47\%$ over the entire range where the decay is 
kinematically allowed, which covers practically the entire parameter range of 
interest; viz. $M_{H_1} <  50 (40)$ GeV for $\Phi_{\rm CP} = 90^\circ 
(60^\circ)$. It can be also seen from both the tables that the 
BR($H^\pm \to H_1 W$) is larger than $90\%$ over most of the parameter space of interest. So not only that $H^+$ can be produced abundantly in the $t$ decay
giving rise to a possible production channel of $H_1$ through the decay $H^\pm
\to H_1 W^\pm$, but this decay mode will be the only decay channel to see
this light ($M_{H^\pm} < M_t)$ $H^\pm$. The traditional decay mode of 
$H^\pm \to \tau \nu$ is suppressed by over an order of magnitude and thus will 
no longer be viable. Thus  the process  
$$
p p \rightarrow \decay{t~~~}{b {\decay{H^+}{{\decay{W}{\ell \nu (q \bar q)}}~~~~~~~~~~{\decay{H_1}{b \bar b}}}}} ~~~~~~~~\hspace{2cm} + \hspace{2cm} \decay{\bar t}{\bar b {\decay{W}{q \bar q(\ell \nu)}}} + ~~~~X
\vspace{4cm}
$$
will allow a probe of both the light $H_1$  {\bf and } a light $H^\pm$ in this
parameter window in the CP-violating MSSM in the CPX scenario.
\begin{figure}[hbt]
\includegraphics*[scale=0.6]{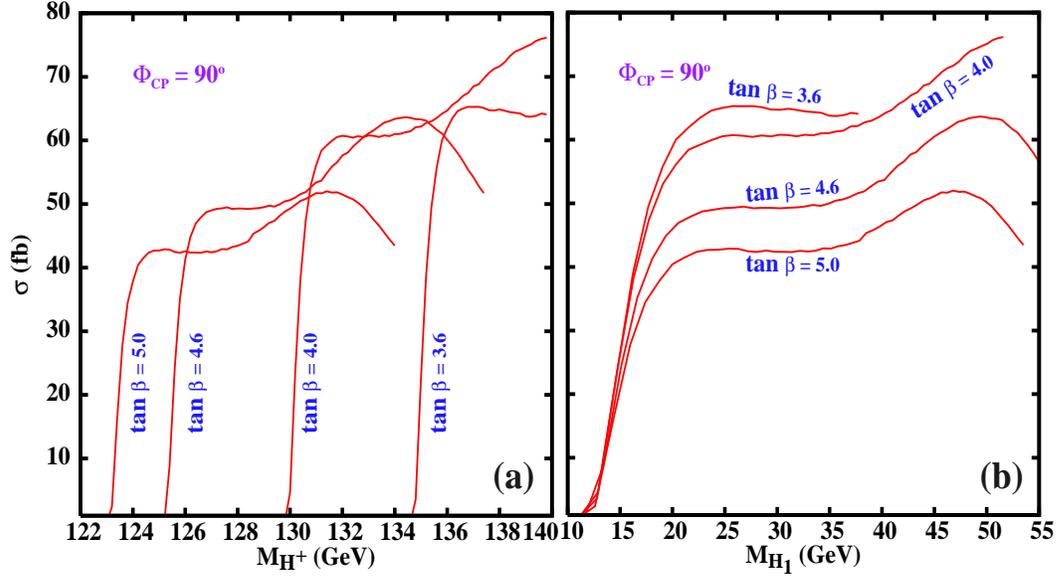}
\caption{Variation of the  cross-section with $M_{H^+}$ $(a)$ and 
$M_{H_1}$ ($b)$ for four values of $\tan\beta =3.6,4,4.6 $ and $5$. 
The CP-violating phase $\Phi_{\rm CP}$ is $ 90^{\circ}$. The other 
MSSM parameters are same as in Figure 1.  These  numbers  should be 
multiplied by $\sim 0.5$ to get the signal cross-section as explained 
in the text. The same mass window cuts as mentioned in Figure 1 have been 
used in this case.}

\label{fig2}
\end{figure}
The signal will consist of three  or more b-tagged and two 
un-tagged jets along with a hard lepton and missing $p_T$.  
For a $b$ tagging efficiency $e$, the suppression factor $SF$ due to the 
demand of three or more tagged $b$ jets is given by

$$
SF  = 4e^3(1-e) + e^4.
$$ 

\noindent Assuming $e=0.5$ we get $5/16$ for this suppression factor.

In our parton-level Monte Carlo analysis we employ following strategies 
to identify final state jets and leptons:\\
\begin{enumerate}
\item $\mid \eta\mid <2.5 $ for all jets and leptons, where $\eta$ denotes 
pseudo-rapidity, 
\item $p_T$ of the hardest three jets to be higher than 30 GeV,
\item $p_T$ of all the other  jets, lepton, as well as  the missing
$p_T$  to  be larger than 20 GeV, 
\item A minimum separation of $\Delta R = 
\sqrt{ \left (\Delta \phi \right)^2 + \left (\Delta \eta \right )^2} = 0.4$.
between the lepton and jets as well as each pair of jets. 
If $\Delta R$ between two partons is less than
$0.4$ we merge them into a single jet. 
\item We impose Gaussian smearing on energies, with 
$\Delta E/E = 0.6/\sqrt{E}$ for jets.
\item We demand three or more tagged $b$-jets in the final state  assuming a 
$b$-tagging efficiency of $50\%$. 
\item The missing $p_T$ is obtained by vector summation of the transverse 
momenta of the lepton and the jets after Gaussian smearing.
\end{enumerate}

Below we  outline the mass reconstruction strategy we employ.
The leptonically decaying W in the above decay chain is reconstructed from 
the lepton momentum  $p_l$ and the missing transverse momentum $p_T$
within a quadratic ambiguity using the constraint that the invariant mass
of the $\ell \nu$ pair $m_{\ell \nu} = M_W$.  In case of complex solutions the 
imaginary part is discarded and the two solutions coalesce. The hadronically
decaying  W is reconstructed from that pair of untagged jets, whose invariant 
mass is 
closest to $M_W$. One top is then reconstructed from one of the reconstructed 
W's and one of the remaining jets chosen such that the invariant mass 
$m_{W jet}$ is closest to $M_t$. Similarly the $H_1$ is then reconstructed 
from a pair from among  the remaining jets, such that the invariant mass 
of the pair is closest to $M_{H_1}$. Then the $H^\pm$  is reconstructed 
from this $H_1$ and the remaining reconstructed W. In case of a quadratic 
ambiguity for the latter, the one giving invariant mass closer to 
$M_{H\pm}$ is chosen. Although the masses of the $H_1$ and $H^\pm$  may not 
be known, one can select the right combinations on the basis of a clustering 
algorithm. Finally the second top is reconstructed by combining this 
$H^\pm$ with the remaining jet.The signal cross-sections shown in 
Figs 1 and 2 are obtained using mass window cuts of 
$M_W \pm 15 $ GeV, $M_t \pm 25 $ GeV, $M_{H_1} \pm 15 $ GeV 
and $M_{H^\pm} \pm 25 $ GeV on the reconstructed 
$W, t, H_1$ and $H^\pm$ masses.  Only the $M_W$ and $M_t$ 
mass window cuts are retained in Figures 3 and 4, 
showing the distributions in the reconstructed $H_1$ and $H^+$ masses.

In Figure 1 we show the variation of the cross-section with $M_{H^+}$ $(a)$ 
and $M_{H_1}$ $(b)$ for the CP-violating phase $\Phi_{\rm CP} = 60^{\circ}$ 
while the choice of other MSSM parameters are defined through Eqs.(11-16). 
We have used the CPSuperH program~\cite{cpsuperh}  with $M_t = 175$ GeV,
to calculate the masses and the couplings of the Higgs-bosons in the CPX 
scenario.  We have used the CTEQ 4L parametrisation of the parton density
distributions and the QCD scale chosen is $2 M_t$.  The numbers presented in 
the figure contain neither the suppression factor due to $b$--tagging 
efficiency nor the $K$--factor(1.3--1.4) due to
the NLO corrections to the $t \bar t $ cross-sections. Taking into account
both, the numbers in the figure should be multiplied by
$5/16 \times 1.3$--$1.4 \sim 0.5$ to get the signal cross-section at the LHC.
It may also be stated that the expected cross-sections at the Tevatron are far
too small for this process to be useful there.

As can be seen from the figure   
the signal cross-section decreases with increase in $\tan \beta$. This
can be explained by the fact that $H^+ \to H_1 W^+$ as well as 
$t \to b H^+$ branching ratio decreases with the increase in $\tan \beta $ for
a fixed $H_1$ mass. 
In this scenario, the largest signal cross-section $(\sim 160$ fb) can be 
obtained for $\tan \beta = 2$ and $M_{H^+} = 135 $ GeV, which corresponds
to $M_{H_1} = 54.3 $ GeV. The cross-section is $\sim 125$ fb for $M_{H^+} 
= 130$ GeV corresponding to $M_{H_1} = 40$ GeV. 
\begin{figure}[hbt]
\includegraphics[scale=0.6]{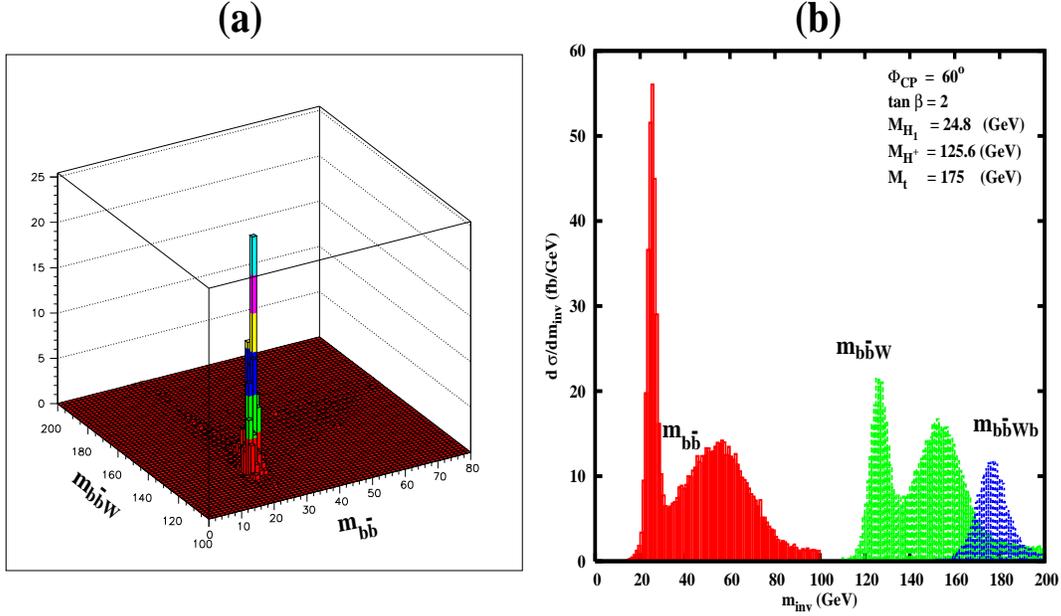}
\caption{Clustering of the $b\bar b, b\bar bW $ and $b\bar b b W$ 
invariant masses: $(a)$ three-dimensional plot for the correlation between
$m_{b\bar b} $ and $m_{b\bar b W} $ invariant mass distribution;
$(b)$ $m_{b\bar b}, m_{b\bar b W}$ and $m_{b\bar b W b}= M_t$
invariant mass distributions for $\Phi_{\rm CP} = 60^{\circ}$.
$M_t, M_W$ mass window cuts have been applied as explained in the text. 
The other MSSM parameters are $\tan\beta = 2, M_{H^+} = 125.6$ GeV and 
the  corresponding light Higgs mass is $M_{H_1} = 24.8 $ GeV.}
\label{fig3}
\end{figure}
In principle there exists a physics background to the signal arising from the
decay $H^\pm \to W^\pm \bar b b$, via the virtual $tb$ channel, but
over this particular range of $M_{H^\pm}$ and $\tan \beta$ the corresponding
branching ratio is negligibly small~\cite{Ma:1997up}.

In Figure 2, we show  
variation of the signal cross-section with $M_{H^+}$ $(a)$ and 
$M_{H_1}$ $(b)$ for the CP-violating phase $\Phi_{\rm CP} = 90^{\circ}$ keeping
other MSSM parameters fixed as in Figure 1.  Apart from the choice
of the phase, the main difference from Figure 1  is in the values of
$\tan\beta$. In this case we have somewhat larger values of $\tan\beta$,
namely $3.6,4.,4.6$ and $5.$, corresponding to the light Higgs window
of Ref.~\cite{carena} for $\Phi_{\rm CP} = 90^\circ$. The largest signal 
cross-section in this case is $\sim 38$ fb.  Note that in both cases the 
signal cross-section is $\gapp  20$  fb for $M_{H_1} \gapp 15$ GeV.

\begin{figure}[hbt]
\includegraphics[scale=0.6]{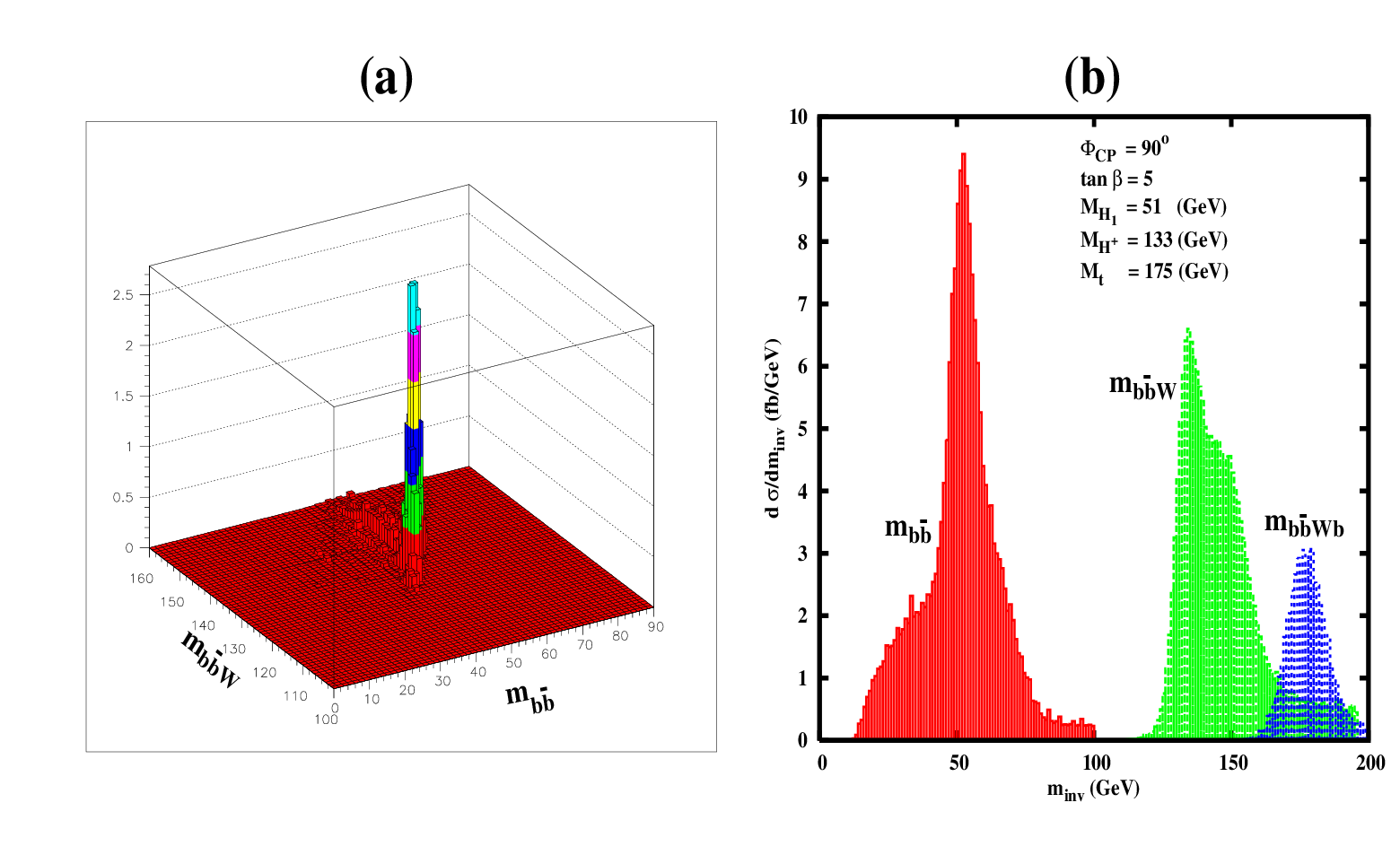}
\caption{Clustering of the $b\bar b, b\bar bW $ and $b\bar b b W$ invariant 
masses. $(a)$ three-dimensional plot for the correlation between 
$m_{b\bar b} $ and $m_{b\bar b W}$ invariant mass distribution. 
$(b)$ $m_{b\bar b}, m_{b\bar b W}$ and $m_{b\bar b W b}= M_t$ 
invariant mass distributions for $\Phi_{\rm CP} = 90^{\circ}$. 
$M_t, M_W$ mass window cuts have been applied as explained in the text. 
The other MSSM parameters are $\tan\beta = 5, M_{H^+} = 133$ GeV, 
corresponding to a light neutral Higgs $H_1$ with mass $M_{H_1} = 51 $ GeV.}
\label{fig4}
\end{figure}
In Figure 3 $(a)$ we show the 
three-dimensional plot for the correlation between $m_{b\bar b}$ and 
$m_{b\bar bW}$ invariant mass distribution for $\Phi_{CP} = 60^{\circ} , 
\tan \beta =2 $ and $M_{H^+} = 125.6$ GeV. The light Higgs mass corresponding 
to this set of input parameter is 24.8 GeV. It is clear from  Figure~\ref{fig3}
that there is simultaneous clustering in the  $m_{b\bar b}$ distribution around
$\simeq M_{H_1}$ and in the $m_{b \bar b W}$ distribution around $M_{H^\pm}$. 
Figure~\ref{fig3}$(b)$ shows the same, in terms of cross-section distribution 
in  $b \bar b$, $b \bar b W$ and  $b \bar b W b$  invariant masses for the 
signal. The clustering feature can be used to distinguish the signal over the 
standard model background. As a matter of fact we estimated the background to
the signal coming from the QCD production of $t \bar t b \bar b$. Even though
the starting LO cross-section  for $t \bar t b \bar b$ production is as high 
as $\sim 8.5$ pb, once all the cuts (including the mass window cuts) are 
applied we are left with a contribution to the signal type events of 
less than  $0.5$ fb.  The major reduction is brought about by requiring that 
the invariant mass of the $b b b W$ be within $25$ GeV of 
$M_t$\footnote{Preliminary studies in ATLAS collaboration presented 
at Les Houches Workshop~\cite{markus_houch} also find that this background can
be suppressed to negligible levels by similar requirements.}.
This makes it very clear that the detectability of 
the signal is controlled primarily by the signal size. It is also clear from  
Figures 1 and 2 that  indeed the signal size is healthy over the regions of 
interest in the parameter space. Thus using this process one can cover the 
region of the parameter space in $\cp$ MSSM, in the $\tan\beta -M_{H_1}$ plane 
which can not be excluded by LEP-2 and where the Tevatron and the LHC have 
no reach via  the usual channels. Note further that this process would be 
the only channel of discovery for the charged Higgs-boson $H^\pm$ as well in 
this scenario, as the traditional decay mode of $H^\pm  \to \nu \tau $ is 
suppressed by over an order of  magnitude. 

Figure~\ref{fig4}$(a)$ shows the three-dimensional plot for the 
correlation between $m_{b\bar b}$ and
$m_{b\bar bW}$ invariant mass distribution for $\Phi_{CP} = 90^{\circ} $, and
somewhat higher values of $\tan\beta$ and $M_{H^+},
\tan \beta =5 $ and $M_{H^+} = 133$ GeV. The light Higgs mass corresponding
to this set of input parameter is 51 GeV. 
The Figure~\ref{fig4}$(b)$ shows the
same, in terms of cross-section distribution in  $b \bar b$, $\bar b b W$
and  $\bar b b b W$  invariant masses for the signal. Both these figures 
show similar clustering of the $b\bar b$, $b\bar b W$ invariant masses 
at values corresponding to $M_{H_1}$ and $M_{H^+}$ respectively as in 
Figure~\ref{fig3}.

It should be mentioned here that the combinatorial background has already been
included in the inclusive $b \bar b$ and $b \bar b W$ invariant mass 
distributions plotted in Figures \ref{fig3}-\ref{fig4} whereas the three 
dimensional plots showing the correlation do not include this. Within the 
framework of the mass reconstruction strategy outlined before, after the
reconstruction of $t \rightarrow bW$, one is left with  
three $b$ jets and a $W$. The former correspond to  three possible invariant 
$b \bar b$ masses for each MC point. It is seen from Figures \ref{fig3} 
and \ref{fig4} that even after inclusion of all the possible pairs at each 
point the peak at the $H_1$ mass is clearly visible.  Now for further 
reconstruction one can choose the pair with invariant mass closest 
to the peak and then calculate the $b \bar b W$ invariant mass by combining 
this pair with the remaining W. In case of quadratic ambiguity for the $W$ 
both the values for the $W b \bar b$ invariant mass  are retained.  Again, 
we see a clear peak at the $H^+$ mass.  Finally combining this with the
remaining $b$ gives the $W b b b$ invariant mass which peaks at $M_t$. 
In case of quadratic ambiguity for the $W$ we have chosen the $Wbb$
combination with invariant  mass closer to the $H^+$ mass peak. 
In the three dimensional  plot of Figures \ref{fig3}-\ref{fig4}
we show the pair of invariant masses corresponding to this combination of
$W bb$ as well as the $b \bar b$ invariant mass closest to $H_1$ mass.  
We have found that about $50\%$ of the signal events will have more than 
one combination of the $b \bar b$ and 
$b \bar b W$ invariant masses in the window $M_{H_1} \pm 15$ GeV and 
$M_{H^+} \pm 25$ GeV respectively, when one includes all the combinations. 
Thus the combinatorial background is important but does not seem to overwhelm 
the signal.

A comment about the  $M_t$ dependence of our results is in order. If the 
value of $M_t$ used is increased from $175$ to $178$ GeV, typically the 
mass difference $M_{H^+}$--$M_{H_1}$  goes up by about $7$--$8$ GeV and thus
the curves 
in Figures 1 and 2 will extend to $M_{H_1}$  values higher by about $7$--$8$ 
GeV.  We, however, have used the more conservative value of $175$ GeV for 
$M_t$.
as the window in the $\tan\beta$-- $M_{H^+}$ window which we explore, 
has  been obtained using $M_t = 175$ GeV in Ref.~\cite{carena}. Since the 
size of the window where LEP has no reach also gets bigger with an increased 
value of $M_t$~\cite{Abbiendi:2004ww,susylim}, the above observation simply 
implies that the region which the process 
$t \to b H^\pm \to b H_1 W \to b b \bar b W$ 
can probe will also be bigger in that case.

\section*{Conclusions}
Thus we have looked in the CPX scenario, in the CP-violating MSSM, at the 
region in the $\tan \beta$--$M_{H^\pm}$ plane, where a light $H_1$ signal 
might have been 
lost at LEP due to strong suppression of the $H_1 Z Z$ coupling and where the
Tevatron and the LHC will have no reach due to a simultaneous suppression of
the $H_1 t \bar t$ coupling as well. Specifically, we concentrated 
in the MSSM parameter space $3.5 < \tan\beta  < 5, M_{H_1} \lapp 50 $ GeV
and $2 < \tan \beta  < 3, M_{H_1} \lapp 40$ GeV, for the common CP violating
phase $\Phi_{\rm CP} = 90^{\circ}$ and $60^{\circ}$ respectively, which 
correspond to the light $H_1$ window of~\cite{carena}.  We find that
a light charged Higgs ($M_{H^\pm} < M_t$) with a large value for 
the branching ratio for the decay $H^\pm \to H_1 W$ is 
realised almost over the 
entire parameter space that we considered. We find that such a light $H_1$ 
and  light $H^\pm$, can be probed at the LHC in $t \bar t$ signal where 
one of the top quarks decays into the $bb\bar b W$ channel, via
$t \to b H^\pm, H^\pm \to W H_1$ and $H_1 \to b \bar b$.  Our parton-level
Monte Carlo yields upto $\sim 1100$--$5000$  events for a ${\cal L}=30$
fb$^{-1}$ corresponding to the CP-violating phase $\Phi_{\rm CP} = 90^{\circ}$
and $60^{\circ}$ respectively.  The events will show a very characteristic
correlation between the $b \bar b$, $b \bar b W$ and $b b \bar b W$ invariant
mass peaks, indicating that the SM background may be negligible.  Further, 
in a considerable part of this region, the branching ratio for the 
$H^\pm \to \tau \nu$ channel, that is normally used for the charged Higgs 
search, is reduced by over an order of magnitude.  Thus, this $t \bar t$
signal will be a probe of {\bf both} a light neutral $H_1$ and a light charged
Higgs $H^\pm$. It is imperative that this investigation is followed up with a 
more exact  simulation using event generator level Monte Carlo and detector 
acceptance effects, which is beyond our means. We hope that the encouraging 
results from this parton level Monte Carlo study will induce the CMS and the 
ATLAS collaborations to undertake such investigations.

\section*{Acknowledgements}
We wish to thank the organisers of the Workshop on High Energy Physics 
Phenomenology 8 (WHEPP8)  in Mumbai, India (Jan. 4-15, 2004) where  
this work was started and the Board for Research in Nuclear Sciences (BRNS) 
in India, for its support to organise the Workshop. The work of DKG is 
supported by the 
Bundesministerium f\"ur Bildung und Forschung Germany, grant 05HT1RDA/6. DKG 
would also like to thank US DOE contract numbers DE-FG03-96ER40969 
for financial support during the initial stages of this work.


\begin{thebibliography}{99}
\bibitem{Eidelman:2004wy}
S.~Eidelman {\it et al.}  [Particle Data Group Collaboration],
Phys.\ Lett.\ B {\bf 592}, 1 (2004), see also {\tt  http://pdg.lbl.gov}.


\bibitem{lep} ALEPH, DELPHI, L3, OPAL, The LEP Higgs Working Group 
for Higgs Boson Searches, Phys. Lett. B {\bf 565} 61, (2003). 
{\tt CERN-EP-2003-011}.

\bibitem{susylim}
LEP SUSY Working Group, {\tt http://lepsusy.web.cern.ch/lepsusy};
LEP Higgs Working Group, {\tt  LHWG-Note 2004-01}.

\bibitem{gunion}
J.~F.~Gunion, B.~Grzadkowski, H.~E.~Haber and J.~Kalinowski, 
Phys. \ Rev. \ Lett. {\bf 79}, 982 (1997), [arXiv:hep-ph/9704410].


\bibitem{Sakharov:dj}
A.~D.~Sakharov,
    Pisma Zh.\ Eksp.\ Teor.\ Fiz.\  {\bf 5}, 32 (1967);  JETP Lett. 6, 24 (1967).

\bibitem{Dolgov:2002kw}
For a recent summary, see A.~D.~Dolgov,
  [arXiv:hep-ph/0211260].

\bibitem{Bennett:2003bz}
C.~L.~Bennett {\it et al.},
    Astrophys.\ J.\ Suppl.\  {\bf 148}, 1 (2003)
    [astro-ph/0302207].

\bibitem{Dine:2003ax}
For a review, see e.g.\
    M.~Dine and A.~Kusenko,
    Rev.\ Mod.\ Phys.\  {\bf 76}, 1 (2004) [arXiv:hep-ph/0303065].

\bibitem{edm1} P. Nath, Phys. \ Rev. \ Lett. {\bf 66}, 2565 (1991);
Y. Kizukuri and N. Oshino, \ Phys. \ Rev. D {\bf 46}, 3025 (1992);
T. Ibrahim and P. Nath Phys. \ Lett. \ B {\bf 418}, 98 (1998), 
[arXiv:hep-ph/9707409]; Phys. \ Rev. D {\bf 57}, 478 (1998), 
[arXiv:hep-ph/9708456]; 
{\it ibid} D {\bf 58}, 019901(E) (1998);
{\it ibid} D {\bf 60}, 079903 (1999); {\it ibid} D {\bf 60}, 119901 (1999);
M. Brhlik, G.J. Good and G.L. Kane, Phys. \ Rev. D {\bf 59}, 115004 (1999)
, [arXiv:hep-ph/9810457];
A. Bartl, T. Gajdosik, W. Porod, P. Stockinger and H. Stremnitzer,
Phys. \ Rev. D {\bf 60}, 073003 (1999), [arXiv:hep-ph/9903402]; 
D. Chang, W.-Y. Keung and
A. Pilaftsis, Phys. \ Rev. Lett. {\bf 82}, 900 (1999), [arXiv:hep-ph/9811202];
S. Pokorski, J. Rosiek and C.A. Savoy, Nucl. 
\ Phys. B {\bf 570}, 81 (2000), [arXiv:hep-ph/9906206];
E. Accomando, R. Arnowitt and B. Dutta, 
\ Phys. \ Rev. D {\bf 61}, 115003 (2000), [arXiv:hep-ph/9907446];
S. Abel, S. Khalil and O. Lebedev, 
\ Nucl. \ Phys. B {\bf 606}, 151 (2001), [arXiv:hep-ph/0103320];
U.Chattopadhyay, T.Ibrahim and D.P.Roy, 
\ Phys.\ Rev.\ D {\bf 64}, 013004 (2001), [arXiv:hep-ph/0012337];
D. A. Demir, M. Pospelov and A. Ritz, \ Phys. \ Rev. \ D {\bf 67}, 
015007 (2003), [arXiv:hep-ph/0208257].

\bibitem{edm2} A. Pilaftsis, \ Nucl. \ Phys. B {\bf 644}, 263 (2002), 
[arXiv:hep-ph/0207277].

\bibitem{edm3} T. Falk, K. A. Olive, M. Pospelov and R. Roiban,
\ Nucl. \ Phys.  B {\bf 560}, 3 (1999), [arXiv:hep-ph/9904393].

\bibitem{aposto1}
A. Pilaftsis, \ Phys. \ Rev. D {\bf 58}, 096010 (1998),[arXiv:hep-ph/9803297]
and Phys. \ Lett. B {\bf 435}, 88 (1998), [arXiv:hep-ph/9805373].

\bibitem{aposto2}
A.~Pilaftsis and C.~E.~Wagner, \ Nucl. \ Phys. \ B {\bf 553}, 3 (1999)
, [arXiv:hep-ph/9902371].

\bibitem{demir} D. A. Demir, \ Phys. \ Rev. D {\bf 60}, 055006 (1999), 
[arXiv:hep-ph/9901389].

\bibitem{jsik}
S.~Y.~Choi, M.~Drees and J.~S.~Lee, 
\ Phys. \ Lett. \ B {\bf 481}, 57 (2000), [arXiv:hep-ph/0002287].

\bibitem{aposto3}M.~Carena, J.~R.~Ellis,
A.~Pilaftsis and C.~E.~Wagner, 
\ Nucl. \ Phys. \ B {\bf 586}, 92 (2000), [arXiv:hep-ph/0003180].

\bibitem{kane-wang}G. L. Kane and L. -T. Wang, Phys. \ Lett. B {\bf 488},
383 (2000), [arXiv:hep-ph/0003198].

\bibitem{carena}
M.~Carena, J.~R.~Ellis, S.~Mrenna,
A.~Pilaftsis and C.~E.~Wagner, 
\ Nucl. \ Phys. \ B {\bf 659}, 145 (2003), [arXiv:hep-ph/0211467].

\bibitem{Abbiendi:2004ww} G.~Abbiendi {\it et al.} [OPAL Collaboration],
{\em Eur. Phys. J. C} {\bf 37},  49 (2004).

\bibitem{cpv-susyhiggs}
A.~Dedes and S.~Moretti, \ Phys. \ Rev.  \ Lett.  {\bf 84}, 22 (2000), 
[arXiv:hep-ph/9908516], \ Nucl. \ Phys. \ B {\bf 576}, 29  (2000), [arXiv:hep-ph/990941];
S. Y. Choi, K. Hagiwara and J.S. Lee, \ Phys. \ Rev. \ D {\bf 64},
032004  (2001), [arXiv:hep-ph/0103294], \ Phys. \ Lett. \ B {\bf 529}, 212
(2002), [arXiv:hep-ph/0110138]; A. G. Akeroyd and  A. Arhrib, 
\ Phys. \ Rev. \ D {\bf 64}, 095018 (2001),[arXiv:hep-ph/0107040];
A. Arhrib, D. K. Ghosh and O. C. W. Kong, Phys. Lett. B {\bf 537}, 217 (2002) 
[arXiv:hep-ph/0112039]; S.~Y.~Choi, M.~Drees, J.~S.~Lee and J.~Song, Eur. Phys. J. 
C {\bf 25}, 307 (2002) [arXiv:hep-ph/0204200].

\bibitem{dpr} M. Drees, M. Guchait and D.P. Roy, 
Phys.\ Lett.\ B {\bf 471}, 39 (1999), [arXiv:hep-ph/9909266].

\bibitem{cpsuperh}
J.~S.~Lee, A.~Pilaftsis, M.~Carena, S.~Y.~Choi, M.~Drees, J.~R.~Ellis and C.~E.~M.~Wagner,
Comput.\ Phys.\ Commun.\  {\bf 156}, 283 (2004)
[arXiv:hep-ph/0307377].

\bibitem{Heinemeyer:2001qd}
S.~Heinemeyer,
Eur.\ Phys.\ J.\ C {\bf 22}, 521 (2001)
[arXiv:hep-ph/0108059].

\bibitem{Ma:1997up}
E.~Ma, D.~P.~Roy and J.~Wudka,
Phys.\ Rev.\ Lett.\  {\bf 80}, 1162 (1998)
[arXiv:hep-ph/9710447].

\bibitem{markus_houch}
Talk presented by Markus Schumacher at the Workshop at TeV colliders at Les 
Houhces, May 1-21, 2005, Summary in the talk by T. Lari, url: {\tt
http://agenda.cern.ch/askArchive.php?base=agenda\&categ=a053250\&id=a053250s10t1
/moreinfo.}
\end{thebibliography}
\end{document}